**Spectroscopic evidence for a molecular orbital Kondo insulator**

**Authors:**

Ke-Jun Xu[1,2]†, Kuan H. Hsu[3,4]†, Nathan Giles-Donovan[1,2]†, Christopher T. Parzyck[4], Gi-Hyeok Lee[5], Wanli Yang[5], Jun Okamoto[6], Hsiao-Yu Huang[6], Di-Jing Huang[6], Joshua J. Kas[7], John Vinson[8], Zhi-Xun Shen[4,9,10,11], Dung-Hai Lee[1,2], Thomas P. Devereaux[3,4,9]*, Wei-Sheng Lee[4]*, and Robert J. Birgeneau[1,2]*

**Affiliations:**

[1] Department of Physics, University of California, Berkeley, California 94720, USA.

[2] Material Sciences Division, Lawrence Berkeley National Laboratory, Berkeley, California 94720, USA.

[3] Department of Materials Science and Engineering, Stanford University, Stanford, California 94305, USA.

[4] Stanford Institute for Materials and Energy Sciences, SLAC National Accelerator Laboratory, 2575 Sand Hill Road, Menlo Park, CA 94025, USA.

[5] The Advanced Light Source, Lawrence Berkeley National Laboratory, Berkeley, CA, 94720, USA.

[6] National Synchrotron Radiation Research Center, Hsinchu 30076, Taiwan.

[7] Department of Physics, University of Washington, Seattle, WA 98195, USA.

[8] Material Measurement Laboratory, National Institute of Standards and Technology (NIST), Gaithersburg, Maryland 20899, USA.

[9] Geballe Laboratory for Advanced Materials, Stanford University, Stanford, California 94305, USA.

[10] Department of Applied Physics, Stanford University, Stanford, California 94305, USA.

[11] Department of Physics, Stanford University, Stanford, California 94305, USA.

† These authors contributed equally to this work.

* Corresponding authors. R. J. B (robertjb@berkeley.edu), W.-S. L. (leews@stanford.edu), T. P. D (tpd@stanford.edu)

**Abstract:**

A Kondo insulator (KI) is a prototypical example of a highly entangled phase of matter, where many-body interactions between local moments and delocalized electrons engender the non-magnetic insulating ground state. Conventionally, the local moments arise from atomic multiplet states with a narrow bandwidth, limiting Kondo coherence to low temperatures. Here, we realize a new paradigm for constructing the KI state with hybridized molecular orbitals in $FeSb_2$. Resonant inelastic X-ray scattering (RIXS) at the Fe *L*-edge reveals distinct signatures of band-like continuum states and localized states. Comparisons with first-principles calculations establish a mixed-configuration ground state with hybridized Fe *d*-Sb *p* molecular orbitals as basis states. By systematically investigating the RIXS momentum, temperature, and doping dependences, we find propagating collective modes commensurate with many-body charge and spin excitations. Our results pave the way for understanding the emerging class of unconventional *d* electron insulators and engineering high temperature Kondo many-body states.

**Main text:**

Two distinct subsystems are required to construct a Kondo insulator: a periodic lattice of local moments and a sea of conduction electrons (*1*). The hybridization of the local moment states and the conduction electrons through the Kondo many-body interaction can result in a gapped ground state (*2*). In conventional 4*f*-electron-based Kondo lattice systems, the local moments are associated with 4*f* multiplet levels, which generally have a small bandwidth much lower than the onsite Coulomb repulsion. These states have strong spin-orbit coupling, and the electronic configuration is governed by Hund's rule and crystal electric field on the local 4*f* atom (*3*). The many-body state in Kondo insulators can host many emergent phenomena including correlation-driven topological physics and quantum oscillations in the absence of a conventional Fermi surface (*4*). Although due to the small bandwidth of the 4*f* states, the Kondo many-body state only gains coherence at low temperatures. More recently, there are ongoing efforts to seek novel Kondo insulators in 5*f* compounds (*5*) and engineered two-dimensional systems (*6*,*7*), with the goals of establishing higher temperature coherent states and tunable platforms. Here, we establish a new paradigm for a generalized way of constructing Kondo insulating states beyond the atomic multiplet picture, where localized hybrid states, also known as molecular orbitals, give rise to the local moments with a larger bandwidth.

We demonstrate this new class of molecular orbital Kondo insulator in a *d*-*p* electron correlated insulator $FeSb_2$. Previous studies of $FeSb_2$ revealed many unusual physical properties. Firstly, the resistivity of $FeSb_2$ is insulating with a low temperature resistivity plateau (*8,9*), reminiscent of that in a prototypical 4*f* Kondo insulator $SmB_6$ (*10*). This is associated with a crossover from bulk-dominated transport to surface-dominated transport (*11*), supported by non-local transport measurements showing that the bulk of $FeSb_2$ is robustly insulating at low temperatures (*12*). Angle-resolved photoemission spectroscopy measurements have also found metallic surface states in $FeSb_2$ (*13*). Magnetic susceptibility shows a non-magnetic ground state with a broad peak above room temperature, consistent with a thermally-induced spin state transition in the framework of Kondo insulators (*14*). However, the associated spin excitation has not been observed in neutron scattering measurements thus far (*15*). Interestingly, the non-magnetic ground state is predicted to be in proximity to an altermagnetic state with non-trivial topology implications (*16*). Furthermore, $FeSb_{2-x}Te_x$ exhibits a low temperature metallic resistivity state at $x < 0.01$, where the charge carrier exhibits heavy electron mass of $\approx 20\ m_e$ (*17*). Upon further doping, ferromagnetism emerges (*18*).

Finally, recent observation of unconventional quantum magneto oscillations at high magnetic fields in $FeSb_2$ (*19*) highlights the most striking similarity to $SmB_6$.

The above physical properties in $FeSb_2$ suggest that its electronic structure is much more complex than that found in theoretical calculations thus far (*20,21,22,23*). Here, we examine this unusual ground state of $FeSb_2$ using resonant inelastic X-ray scattering (RIXS). The sensitivity of the RIXS process to inter-orbital transitions allows us to probe the multiplet-like features of the system, while the momentum resolution enables the characterization of propagating degrees of freedom linked to itineracy. RIXS can also couple to collective excitations which may be born out of the many-body state (*24*).

**X-ray absorption and inelastic scattering profiles**

The crystal structure of $FeSb_2$ and the experimental geometry are shown in Fig. 1A. We first characterize the XAS spectrum of $FeSb_2$ at 30 K in Fig. 1B, taken with total fluorescence yield (TFY), and compare it with the calculated XAS spectrum using the OCEAN package (*25,26*) (See methods and supplementary materials). The $FeSb_2$ Fe $L_3$-edge XAS spectrum is distinct from the typical ionic $Fe^{2+}$ or $Fe^{3+}$ states (see Ref. *27* and Fig. S1), suggesting that the electronic configuration is complicated. Previous works typically assume local atomic orbitals, where the tilted Sb octahedra nominally splits the lower energy $t_{2g}$ levels into Λ levels of $d_{yz}$ and $d_{xz}$ character and a Ξ level of $d_{xy}$ character. Pioneering work by Goodenough argues that $FeSb_2$ is a $d^4$ system with filled Λ levels (*28*). However, a recent XAS work on $FeSb_2$ argued that the $d^4$ configuration is inconsistent with the XAS results and instead suggested a $d^6$ model with a temperature-induced transition between $Fe^{2+}$ low spin and $Fe^{2+}$ high spin states (*29*). However, we find good agreement between the experimental XAS spectra and that obtained from OCEAN (Fig. 1B), which is based on density-functional theory and the Bethe-Salpeter equation (*25,26*). This implies that the XAS signal is dominated by continuum-like states in the band structure framework rather than the localized ionic picture, despite the insulating ground state.

We expand on the investigation of the electronic configuration by examining the RIXS spectra. We first perform RIXS measurements in the *b-c* scattering plane with the *b* axis oriented to the normal of the sample surface (Fig. 1A). Fig. 1C shows the incidence-energy-dependent RIXS distribution curves from 704.8 eV to 708.2 eV in 0.2 eV steps. At incident energies around 706 eV, we observe Raman-like features around 200 meV (M1, red dots) and 800 meV (M2, blue dots). The M1 mode is observed with both σ (E // *a* axis) and π (E mostly along the *b* axis) incident photons (Fig. S2), and it is slightly more prominent with π photons. At higher incidence energies around 707.5 eV, we observe additional energy loss features around 250 meV (M3, black circles) and 600 meV (M4, black dots). We compare the experimental RIXS spectrum (Fig. 1D) with those calculated using two different methods: OCEAN calculations using a GW-corrected band structure (Fig. 1E), and the charge transfer Hamiltonian, full atomic multiplet (CTHFAM) model (*30*) (Fig. 1F). Here, the former considers the continuum states within the band structure picture, whereas the latter considers the hybridized multiplet states. The spectrum above 1 eV energy loss is dominated by fluorescence-like emission, showing good agreement with the OCEAN RIXS spectrum. The highest intensity in the experimental RIXS spectrum at around 707.8 eV is consistent with the XAS resonance. The Raman-like features below 1 eV energy loss (red arrow in Fig. 1D) are not captured by the OCEAN RIXS calculations (Fig. 1E and Fig. S3), but instead are consistent with the low energy features in the CTHFAM spectrum (red arrow in Fig. 1F).

To highlight the behavior of the Raman-like modes, we can examine their resonance profiles. Fig. 1F shows the integrated RIXS intensities between 0.06 eV to 0.6 eV and between 0.06 eV and 7 eV. The larger window encompasses both orbital features and the fluorescence feature, and indeed the profile looks similar to the XAS. The small window highlights the low energy Raman-like features, showing a different line shape than the TFY XAS. The α feature near 706 eV is enhanced and the β feature, originally near the TFY XAS resonance, is much weaker. We do not observe additional features at incident energies above 708 eV (Fig. S4).

To check the validity of the CTHFAM spectrum regarding the low energy features, we have also performed a systematic search of the electronic configuration space with multiplet calculations with a single transition metal ion (Fig. S5). Importantly, the $d^6$ S=0 state (low spin $Fe^{2+}$) does not host low energy excitations, in contrast with the experimental results. Compared to the experimental RIXS spectrum, the CTHFAM spectrum indeed matches the best out of all the calculated multiplet spectra. The CTHFAM state is a mixed-configuration ground state with main contributions from $|d^6\underline{L}^4\rangle$, $|d^7\underline{L}^5\rangle$, $|d^8\underline{L}^6\rangle$, and a minor contribution from $|d^5\underline{L}^3\rangle$ (see table S2). This hybridized state does not have well-defined spin quantum numbers due to spin-orbit coupling, but is associated with an S=1 ground state when the spin-orbit coupling is turned off. The proximity to an effective intermediate spin state suggests that holes on the Sb ligand can partially compensate the Fe *d* electron moments, resulting in a new pseudospin in the mixed-configuration wave function basis. In this sense, the effective local moment in $FeSb_2$ is not based on purely *d* electrons, but rather born out of the Fe-Sb molecular orbitals (Fig. 1H-I). The non-zero moment observed in the calculated ground state suggests that Kondo screening is required to satisfy the experimentally observed non-magnetic ground state.

**Phenomenology of low energy excitations**

We focus on the M1 and M2 modes, which are present at the absorption edge and most relevant to the valence states. We explore their momentum structures with an incidence photon energy of 705.8 eV, where the M1 feature is most prominent. Fig. 2A shows the momentum space trajectories in the *b*-*c* scattering plane that we have undertaken. Cut 1 roughly follows the *b* axis and is obtained by performing θ-2θ scans, with θ being close to the specular reflection position. We find negligible dispersion of the M1 and M2 features along the *k* direction (*b* axis) within this measured momentum range (Fig. 2B). Cut 2, which follows a circle in reciprocal space, is obtained by fixing 2θ at 150 degrees and rotating the sample θ angle. Due to the lack of dispersions of the M1 and M2 features along the *b* axis, we can project cut 2 onto the *c* axis and attribute the dispersion to the *l* direction. A few representative curves and fits at select θ angles are shown in Fig. 2C, and the extracted dispersions for M1 and M2 are shown in Fig. 2D. We also perform momentum dependence measurements in the *a*-*b* scattering plane. Representative RIXS curves and fits at different θ angles in the *a*-*b* plane are shown in Fig. 2E, with the projected dispersion along *h* shown in Fig. 2F. Remarkably, we observe that M2 only exhibits dispersion along the *c* axis, meaning it behaves like a one-dimensional collective mode. On the other hand, M1 is dispersive along *a* and *c*, but is apparently dispersionless along *b*, indicating that it behaves as a two-dimensional collective mode. The full set of momentum-dependent data is shown in Fig. S6.

We can gather more information on the nature of M1 and M2 by varying the temperature. Fig. 3A shows a comparison between the incidence-energy-dependent RIXS spectrum at 30 K and at 300 K. M3, M4, and the fluorescence-like features at energies above 1 eV all remain largely unchanged throughout this temperature range. Remarkably, the M2 feature seems to evolve from Raman-like at low temperatures to fluorescence-like at high temperatures. If we examine the temperature evolution of the low energy resonance profile (inset of Fig. 3A), we can see that the β feature corresponding to M3 and M4 remains constant, while the α feature shifts to lower energy due to the changes in M2. We quantitatively examine the temperature evolution of M1 and M2 by fitting the spectra at different temperatures (Fig. 3B). The results are summarized in Fig. 3C-F. The energy position of M1 remains approximately the same between 30 K and 300 K, but the width broadens with increasing temperature. For M2, the energy position decreases with temperature at a rate of –0.61 meV/K for an incidence energy of 705.8 eV, while the width stays approximately the same.

Another useful strategy for probing the nature of the low energy excitations is to perturb the ground states via chemical substitution and then observe the subsequent change of the excitation spectrum. Here we systematically examine the RIXS spectra of $FeSb_{2-x}Te_x$ for x = 0.02 to x = 0.5. Fig. 4A-C shows incident-energy-dependent RIXS spectra for three representative dopings. We observe that the M1 feature is suppressed with increasing doping. This can be more clearly seen in the raw RIXS data in Fig. 4D. Here the data shown is at θ = 10° and 2θ = 150°, where the M1 feature is sharpest and well separated from M2. We observe that the M1 intensity is suppressed with doping, while the energy scale slightly hardens. At the same time, the energy and intensity of the M2 feature remains largely unchanged. To ensure that we have not missed the M1 feature due to any dispersions of this feature, we also show the momentum dependent measurements of the x = 0.4 sample in Fig. 4E. Here we see that the M1 mode is suppressed throughout this momentum range.

**Propagating many-body excitations**

We first discuss the behavior of the M2 mode in the RIXS spectra. Due to its energy scale of around 800 meV, we assign this feature to an inter-orbital excitation. Because the ground state has a mixed-orbital configuration, the inter-orbital excitations are not conventional *d-d* excitations but rather *dp-dp** excitations from one molecular orbital to another. Examining the projected orbital content of the ~800 meV mode in the CTHFAM model (Table S4), we find that this excitation involves significant electron weight transfer from Fe *d* orbitals to Sb *p* orbitals, consistent with a charge type molecular orbital antibonding excitation. The large dispersion of M2 indicates that it can propagate along the *c* axis and form collective one-dimensional excitations. We note that the *c* axis in $FeSb_2$ is substantially smaller than the *a* and *b* axes, which may result in a much larger orbital overlap along the *c* axis. The large dispersion is consistent with the expectations that the Sb *p* orbital components in the molecular orbitals can be extended in real space. These itinerant orbital excitations are phenomenologically similar to the orbitons observed in certain one-dimensional and two-dimensional cuprates (*31,32*), as well as the perovskite iridate $Sr_2IrO_4$ (*33*). Another unusual behavior of the M2 mode is the temperature-induced change from being Raman-like at low temperatures to being fluorescence-like at high temperatures (Fig. 3A), which occurs smoothly and in a temperature range without a structural transition. We can understand this by considering the single particle spectrum as a function of temperature in a Kondo system. In this case, the localized states with a small bandwidth are broadened significantly at high temperatures by strong spin scattering. The continuous crossover from local atomic-like excitations to continuous excitations observed here in $FeSb_2$ is compatible with such a Kondo lattice model.

We now consider origins of the M1 mode. Its energy scale, which is around 200 meV to 300 meV, is too high for phonons, whose highest energy is about 37 meV (*34,35*). Magnons (*36*) and plasmons (*37*) are nominally forbidden in the non-magnetic insulating ground state of $FeSb_2$. M1 is also not likely a bandgap exciton, as the lower bound of the electron-hole continuum typically does not form a sharp peak as is observed in $FeSb_2$, but rather forms an edge with intensity suppression within the band gap (*38*). In addition, if we extrapolate the absence of dispersion along the *b* axis to the zone center, the zero-momentum-transfer mode energy of 300 meV is considerably larger than the optical gap of around 75 meV to 130 meV (*39*). In fact, optical measurements did not observe any distinct features in the energy range of 200 meV to 300 meV (*40*), suggesting that M1 is optically dark. We note that the dispersions of M1 and M2 along the *h* direction are quite different, suggesting that the origin of these modes are distinct. Furthermore, Te doping induces a suppression of the M1 intensity while the M2 intensity remains largely unchanged, further supporting the case that M1 has a different origin than M2.

We postulate that a candidate for M1 is a pseudospin excitation analogous to the spin excitation from the singlet Kondo ground state to a triplet excited state (*41*), which has been observed in $SmB_6$ with inelastic neutron scattering (*42,43,44*). Considering the non-magnetic ground state, previous signatures of Kondo physics in $FeSb_2$ (*14*), and the close proximity to ferromagnetism through various dopants (*18,45,46*), the singlet-triplet excitation is an attractive explanation for the M1 mode. In support of this scenario, the projected orbital content of low energy excitation in the CTHFAM calculations (Table S4) shows a weight transfer character primarily between Fe *d* orbitals, which would be consistent with a moment reversal type of molecular orbital excitation without charge redistribution. Upon Te doping into a more metallic state, the spin gap closes and this pseudospin excitation would become unobservable, consistent with the observations in Fig. 4. Because of the relatively large energy scale of M1, previous inelastic neutron scattering (INS) measurements on $FeSb_2$ (*15*) using a low incident neutron energy could not observe this mode. We expect that an INS experiment with a high incident energy can resolve this excitation and map out the form factor.

**Dual nature of electronic states in $FeSb_2$**

Our systematic RIXS investigations, combined with theoretical calculations, reveal the dual nature of the electronic states in $FeSb_2$. On the one hand, our CTHFAM model shows that the ground state contains mixed-configuration electronic states, forming molecular orbitals comprised of Fe *d* and Sb *p* states. However, the itinerant aspect of $FeSb_2$ is better captured by OCEAN calculations derived from band structure. Microscopically, the combined effects of Hund's coupling, crystal-field splitting, and anisotropic hybridization can result in different degrees of itinerancy for different orbitals. In $FeSb_2$ with 2 distinct Fe sites, such effects could lead to separation of the Fe 3*d* manifold into orbitals that are more localized and others that are more strongly hybridized with Sb *p* states and therefore more itinerant. In this context, the more localized molecular orbitals would play a role analogous to the localized *f* orbitals in canonical Kondo systems, while the more itinerant molecular orbitals form conduction-like bands (Fig. 1I). It is the combination of these two aspects of the *d-p* orbitals that ultimately forms a molecular orbital Kondo insulator in $FeSb_2$.

In the conventional Kondo insulator $SmB_6$, the ligand B has electronic states far away from the Fermi level, and the local moment and itinerant electrons are comprised of Sm *f* electrons and Sm *d* electrons respectively. Whereas in $FeSb_2$, the ligand Sb *p* states play an active role in the wave functions of the ground state and low energy excitations. The calculated CTHFAM ground state implies that $FeSb_2$ has a much more complex mixed-valent character compared to the 4f compounds, though exhibiting similar types of antibonding and spin flip types of excitations (*41*). The larger energy scales and higher temperature scales in $FeSb_2$ compared to conventional 4*f*

Kondo systems enable a much larger range of experimental techniques for probing the unconventional ground state, and opens a new paradigm for engineering high temperature molecular orbital Kondo insulators. We expect that this picture may be applicable to a number of other Fe-based correlated insulators with even higher temperature scales, including FeSi (*47*), $FeGa_3$ (*48*), and $Fe_2VAl$ (*49*).

**Acknowledgements**

The authors thank Suchitra Sebastian, Nicholas Popiel, Brian Moritz, and Chunjing Jia for insightful discussions, as well as Ganesh Channagoudra and Cheng-Chi Chu for assistance with the RIXS experiments. The authors are also grateful to James Analytis, Valeria Rosa Rocha, and Yuanqi Lyu for checking the sample compositions with energy-dispersive spectroscopy.

**Funding:**

U.S. Department of Energy, Office of Science, Office of Basic Energy Sciences, Materials Sciences and Engineering Division, Contract No. DE-AC02-05-CH11231 within the Quantum Materials Program (KC2202) (RJB)

National Synchrotron Radiation Research Center (NSRRC), Hsinchu, Taiwan, proposal number 2024-2-251 (KJX, NGD, WSL, RJB)

Advanced Light Source (ALS), a DOE Office of Science User Facility, contract no. DE-AC02-05CH11231 (KJX, GHL, WY, RJB)

Sherlock cluster at Stanford University (KHH, TPD)

National Energy Research Scientific Computing Center (NERSC), a Department of Energy Office of Science User Facility, NERSC award BES-ERCAP0027203 (KHH, TPD)

**Author contributions**

Conceptualization: KJX, ZXS, WSL, RJB

Methodology: GHL, WY, JO, HYH, DJH, KHH, JJK, JV, TPD

Investigation: KJX, NGD, CTP, WSL, KHH, DHL

Visualization: KJX, KHH

Funding acquisition: TPD, RJB

Project administration: KJX, TPD, WSL, RJB

Supervision: TPD, WSL, RJB

Writing – original draft: KJX, KHH

Writing – review & editing: KJX, KHH, DHL, TPD, WSL, RJB

**Competing interest declaration**

Certain equipment, instruments, software, or materials are identified in this paper in order to specify the experimental procedure adequately. Such identification is not intended to imply recommendation or endorsement of any product or service by National Institute of Standards and Technology (NIST), nor is it intended to imply that the materials or equipment identified are necessarily the best available for the purpose.

**Data, code, and material availability:**

Code used to generate computational data can be accessed at doi.org/10.5281/zenodo.15659552. Source data will be available at the Dryad repository (URL to be updated). Samples used for this study are available upon request. Details for reproducing the sample synthesis are provided in the supplementary materials.

**Supplementary Materials**

Materials and Methods

Supplementary Text

Figs. S1 to S8

Tables S1 to S4

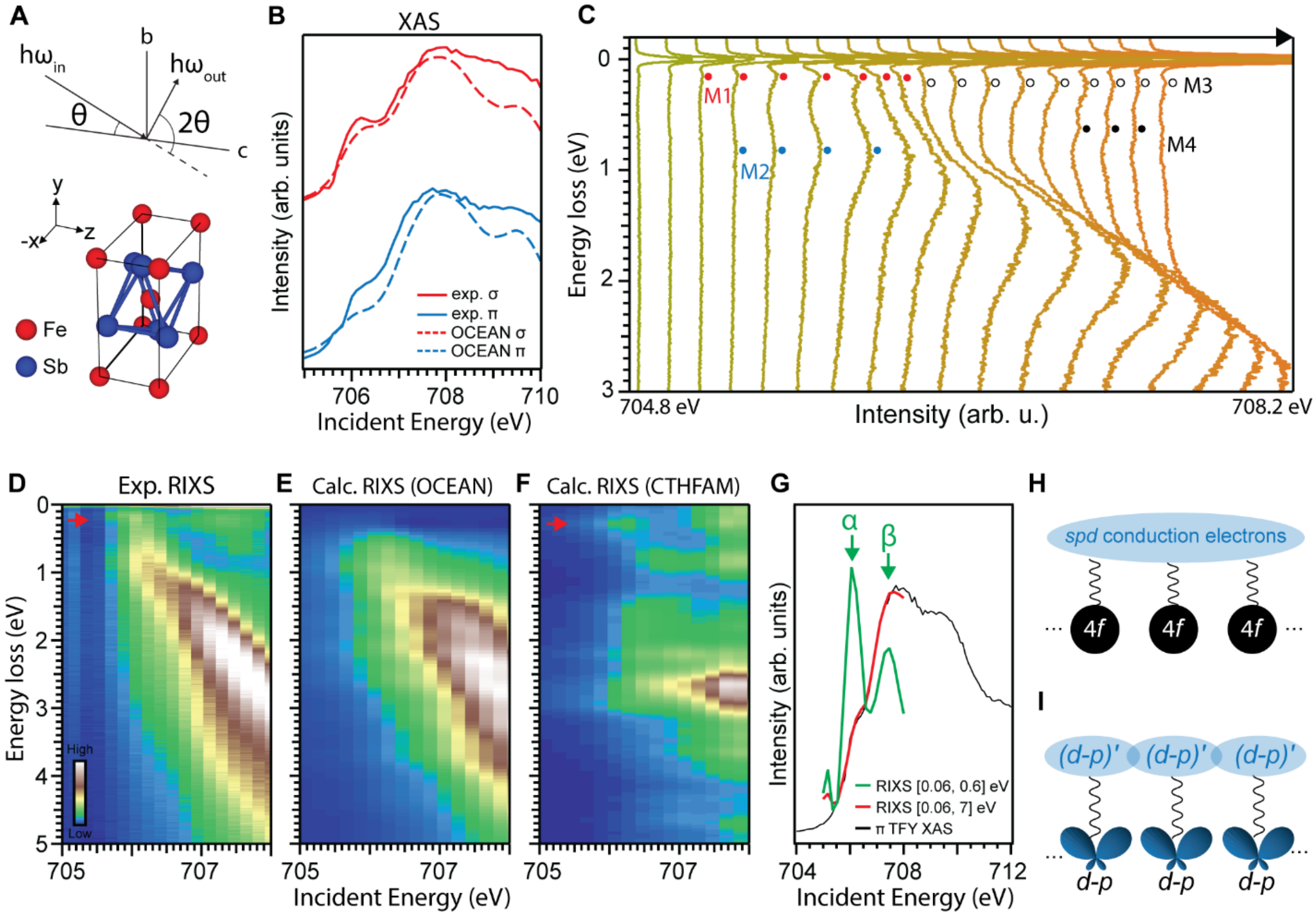

**Fig. 1 |** X-ray spectroscopy reveals molecular orbital states in $FeSb_2$. (A) Crystal structure of $FeSb_2$ and the scattering experiment geometry. θ is defined as the angle between the incident beam and the sample surface. 2θ is defined as the angle between the incident beam and the scattered beam。(B) Experimental (solid lines) and calculated (dashed lines) XAS spectra of $FeSb_2$ at the Fe $L_3$-edge. Experimental spectrum is taken at a grazing incidence geometry with total fluorescence yield (TFY). σ polarization indicates electric field in the vertical direction perpendicular to the scattering plane (*a* axis) and π polarization indicates electric field within the horizontal scattering plane (mostly along *b* axis at grazing geometry). (C) Distribution curves of the incident-energy-dependent RIXS spectrum from 704.8 eV to 708.2 eV in 0.2 eV steps. There are at least 4 Raman-like features: M1 (red dots), M2 (blue dots), M3 (black open circles), and M4 (black dots). (D) Experimental incidence-energy-dependent RIXS spectrum color plot. (E) Calculated incidence-energy-dependent RIXS spectrum using OCEAN code that incorporates GW corrections (see methods and supplementary materials). (F) RIXS spectrum calculated using a charge transfer Hamiltonian, full atomic multiplet (CTHFAM) formalism that includes hybridization with Sb atoms (see methods). Red arrow in (D) and (F) highlight the low energy Raman-like excitation. (G) Resonant profile of the Raman-like modes and comparison to TFY XAS. Integrated RIXS intensity in the energy loss window between [0.06, 0.6] eV (green curve) and between [0.06, 7] eV (red curve). The black curve is the grazing geometry TFY XAS with π polarized photons. Labels "α" and "β" correspond to the features in the green curve. The experimental data here are taken at 30 K. (H) Schematic of a conventional atomic Kondo lattice system with local 4f states interacting with a bath of conduction electrons. (I) Schematic of a molecular orbital Kondo lattice where local moments are born from hybrid *d-p* states.

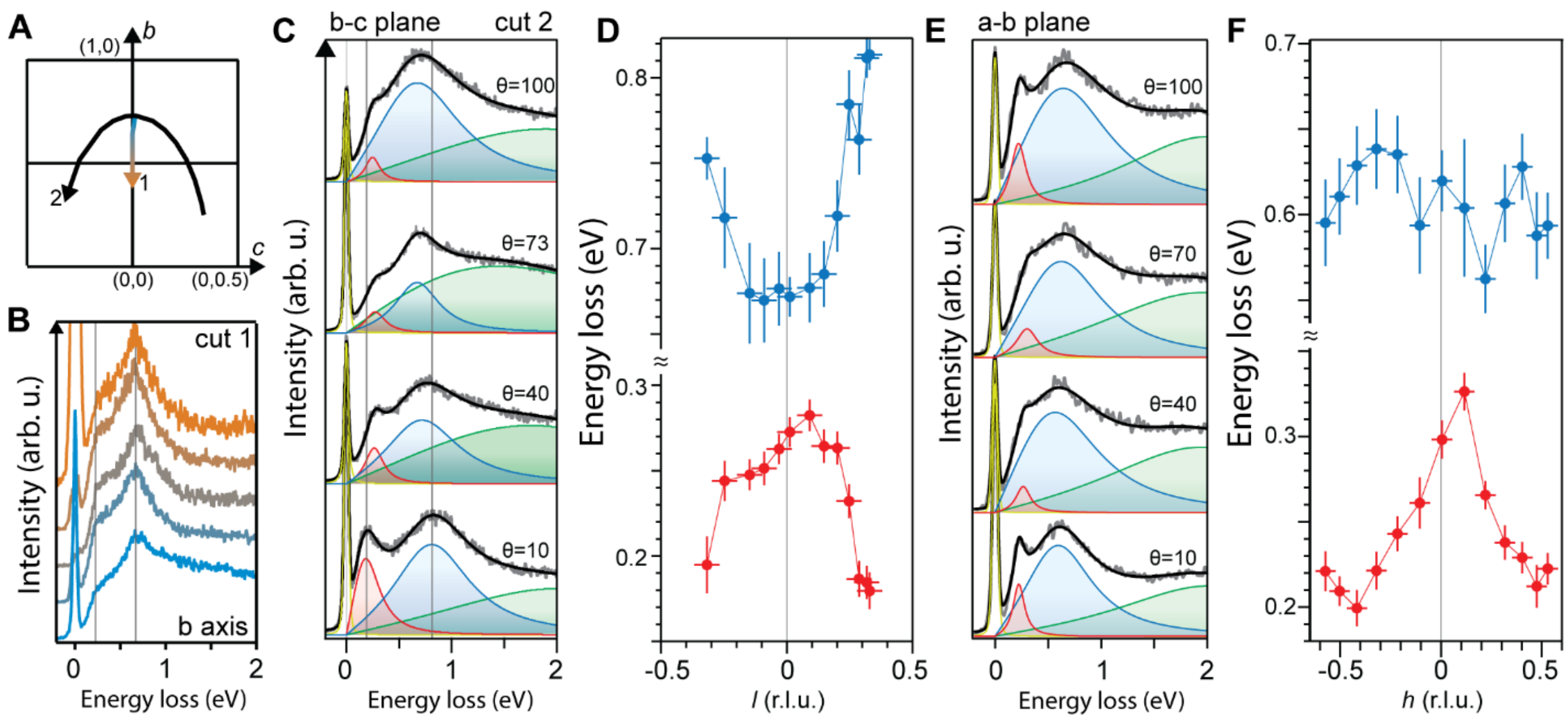

**Fig. 2 |** Dispersive low energy excitations in three-dimensional momentum space. (A) Schematic drawing of the spectral cuts in the *b*-*c* plane of the reciprocal space. (B) RIXS curves along cut 1, which is approximately along the *b* axis. (C) Select RIXS curves at different θ angles with a fixed 2θ = 150° in the *b*-*c* scattering plane, which follows the trajectory in cut 2 shown in (A). The fit consists of a quasielastic peak (yellow), M1 (red), M2 (blue), and a high energy charge transfer spectral weight (green). The quasielastic peak is fitted with a Voigt function, and the other features are fitted with antilorentzians. The peaks were convolved with a gaussian of the quasielastic peak width in the fit. (D) Extracted dispersions of M1 (red) and M2 (blue) projected onto the *l* direction (*c* axis). The peak position data points are obtained by finding the maximum of the fitted antilorentzians. The vertical error bars are the 1 sigma fit confidence interval. (E) Select RIXS spectra at different θ angles with a fixed 2θ = 150° in the a-b scattering plane. The fitting is performed in the same way as in (C). (F) Extracted dispersions of M1 (red) and M2 (blue) projected onto the *h* direction (*a* axis). The peak position data points are obtained by finding the maximum of the fitted antilorentzians. For all spectra here, incident photon polarization is π and the temperature is 30 K.

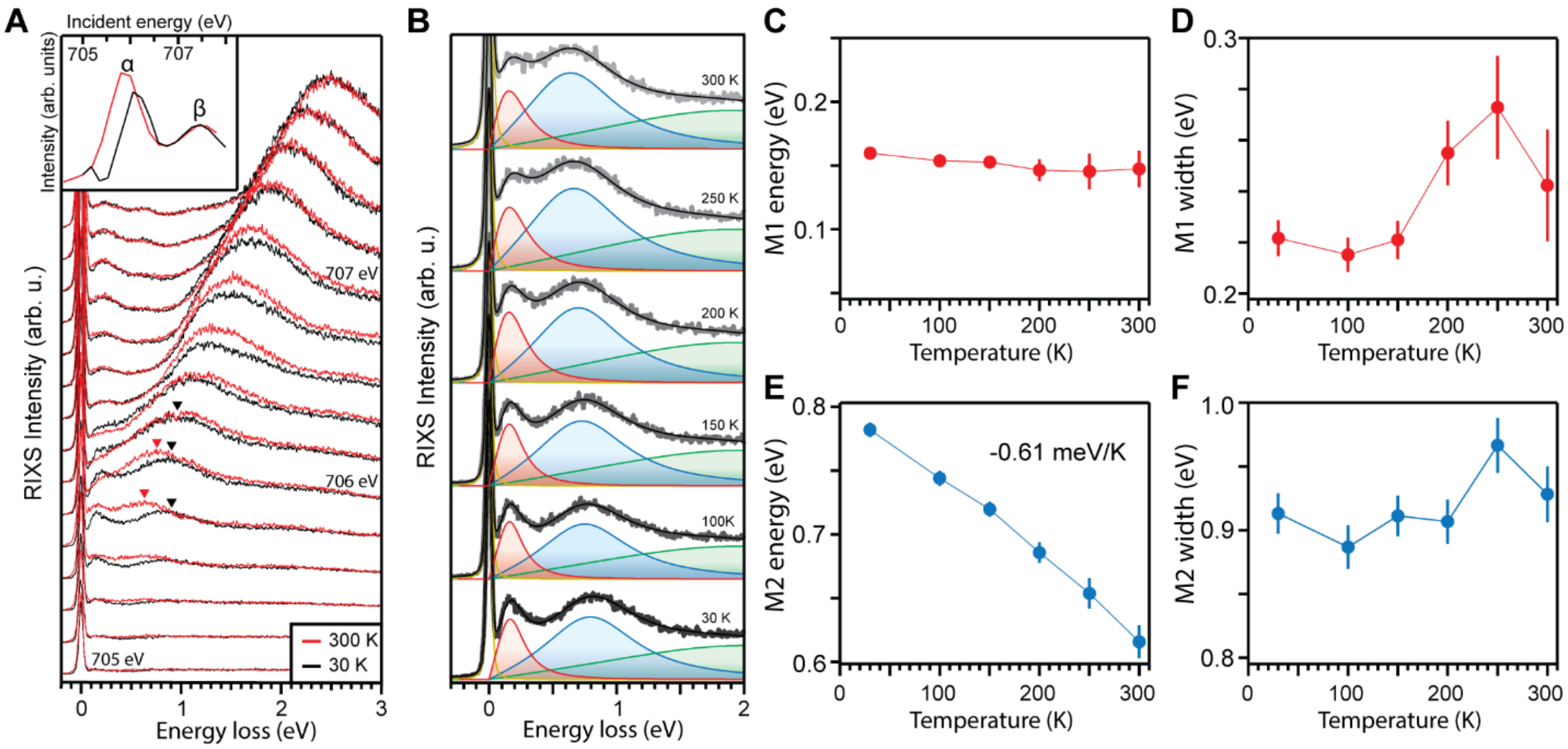

**Fig. 3 |** Unusual temperature-dependent evolution of low energy excitations. (A) Incident-energy-dependent RIXS spectra at 30 K (black curves) and 300 K (red curves) at θ = 10° and 2θ = 150° in the *b-c* scattering plane. Triangles highlight the different behaviors of the M2 mode at low and high temperatures. Inset shows the resonant profile of low energy features integrated between [0.06, 0.6] eV at 30 K (black curve) and 300 K (red curve). (B) Temperature-dependent RIXS curves at 705.8 eV incidence photon energy in the same geometry as (A). The fits are performed in the same way as described in Fig. 1. (C-F) Extracted fit parameters for M1 peak position (C), M1 peak width (D), M2 peak position (E), and M2 peak width (F), as a function of temperature. The peak position (width) data points are obtained by finding the maximum (full-width-half-max) of the fitted antilorentzians. The vertical error bars are the 1 sigma fit confidence interval. For all spectra here, incident photon polarization is π and the temperature is 30 K.

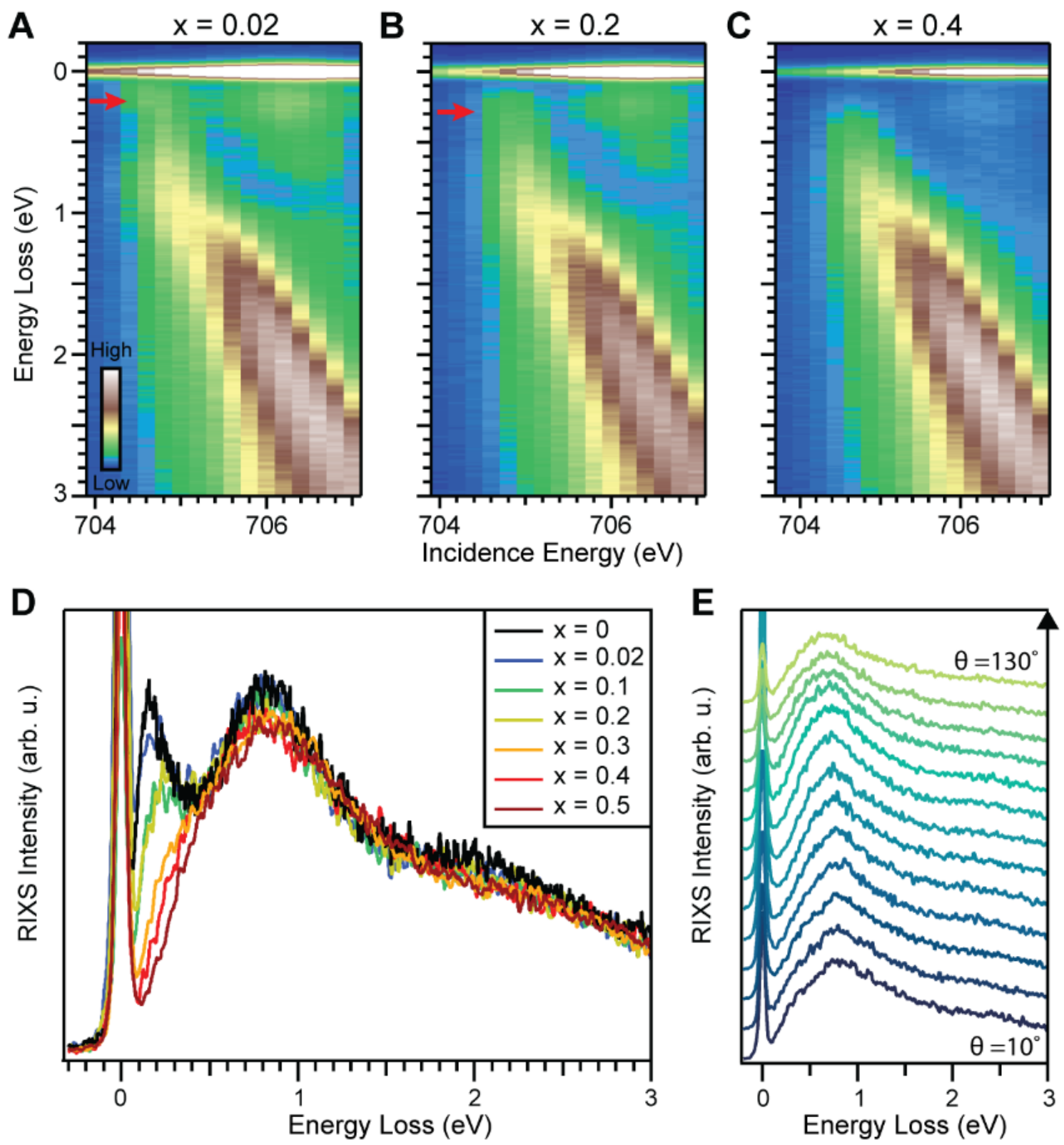

**Fig. 4 |** Doping dependence of low energy excitations. (A-C) Incident-energy RIXS maps for $FeSb_{2-x}Te_x$ samples with x = 0.02 (A), x = 0.2 (B), and x = 0.4 (C). Red arrows point to the evolution of the M1 feature with increasing Te doping. (D) RIXS curves for different dopings at θ = 10° and 2θ = 150°. The curves are normalized using the portion of the spectrum above 2 eV, and is only used for qualitative comparisons. (E) Momentum dependence of $FeSb_{1.6}Te_{0.4}$ in the *a-c* scattering plane, same as cut 2 in Fig. 2a. Curves are from θ = 10° to θ = 130° in 10° steps, while 2θ is fixed at 150°. Incident photon polarization is π.

# Supplementary Materials for

**Spectroscopic evidence for a molecular orbital Kondo insulator**

Ke-Jun Xu, Kuan H. Hsu, Nathan Giles-Donovan, Christopher T. Parzyck, Gi-Hyeok Lee, Wanli Yang, Jun Okamoto, Hsiao-Yu Huang, Di-Jing Huang, Joshua J. Kas, John Vinson, Zhi-Xun Shen, Dung-Hai Lee, Thomas P. Devereaux*, Wei-Sheng Lee*, and Robert J. Birgeneau*

Corresponding authors: R. J. B (robertjb@berkeley.edu), W.-S. L. (leews@stanford.edu), T. P. D (tpd@stanford.edu)

**The PDF file includes:**

Materials and Methods
Supplementary Text
Figs. S1 to S8
Tables S1 to S4

## Materials and Methods

Sample synthesis and preparation

The $FeSb_2$ crystals were synthesized with a conventional self-flux method. Due to the incongruent melting nature of $FeSb_2$, the growth must occur with excess Sb. Sb shot (99.9999% by at.) were mixed with Fe pieces (99.99% by at.) in a 9:1 atomic ratio. The mixed charges were sealed inside quartz tubes along with a quartz wool filter. The tubes were heated up to 850°C and held for 5 hours to homogenize before cooling to 780°C in 5 hours. The crystal growth occurs from 780°C to 650°C at a cooling rate of 0.78°C per hour. Following the growth, the excess Sb were removed through the quartz wool filter by centrifuge.

For the XAS and RIXS measurements, we first prepared a flat surface in the *a-c* plane by cleaving the crystals. It is known from previous ARPES experiments that the *a-c* plane surface is a natural cleavage plane. We then placed the flat cleaved surface downwards and glue the samples to a 5 mm by 5 mm aluminum plate using EPO-TEK H20E silver epoxy for grounding and Loctite EA1C epoxy (also known as Torrseal) for strength. We oriented the in-plane direction and the surface normal by using Laue back scattering, before securing a ceramic top post for cleaving. We cleaved the crystals again immediately before the X-ray measurements to minimize the exposure time of the surface to air.

RIXS measurements

We performed the high resolution RIXS measurements at Beamline 41A of the Taiwan Photon Source using the active grating monochromator (AGM) and active grating spectrometer (AGS) scheme. The details of the AGM-AGS experimental end station can be found elsewhere (*50*). We performed the X-ray measurements near the Fe $L_{2,3}$-edge in the photon energy range 700 eV to 730 eV. The resolution of the spectrometer was optimized using the elastic peak scattered from a rough Carbon tape surface that was mounted adjacent to each sample, and is about 40 meV. The sample temperature during the measurements was 30 K except during the temperature-dependent measurements. The scattering plane, defined by the rotation of sample θ and scattering arm 2θ, is horizontal. For measurements in the *b-c* plane, the *c* axis was oriented horizontally with the surface normal being the *b* axis. For measurements in the *a-b* plane, the *a* axis was oriented horizontally with the surface normal being the *b* axis. For the measurements along the b axis, the spectrometer arm was continuously swept through the 2θ angle without breaking the vacuum. Preliminary room temperature XAS and low energy resolution RIXS experiments were carried out at beamline 8.0.1 of the Advanced Light Source, using the iRIXS instrument (*51*) with a fixed 90° scattering angle and 45° incidence angle.

## Supplementary Text

First-principles calculations of XAS and RIXS spectra

All DFT ground state calculations were done using the Quantum Espresso DFT package (*52*). We generated pseudopotentials using the ONCVPSP code (*53*) (version 3.3.1) with PBE parametrization. Full cell relaxation with symmetry constraint of the space group *Pnnm* is performed with the generalized gradient approximation (GGA), without the consideration of spin-polarization effect, as noted in the prior report (*22*). The force convergence criterion is set to $<10^4$ Ry/Bohr for all force components with a plane-wave cutoff of at least 160 Ry. The ground state calculations are performed using a gamma-centered k grid with a linear density of at least 0.2 $Bohr^{-1}$ using Gaussian smearing with σ = 0.02 eV. Projected density of states (pDOS, Fig. S7), band structures (Fig. S8) and Bader charge (*54*) (Table S1) are calculated based on the converged wavefunctions.

The OCEAN code (version 3.2.2) (*24, 25*) was used to calculate Fe $L_3$-edge XAS and RIXS. The same pseudopotentials and wavefunctions as the ground state calculation are used for the OCEAN calculation. A scissor correction of 0.18 eV was used as an approximate *GW* correction, opening up a gap of 70 meV to match experimental reports of $FeSb_2$ as a narrow gap semiconductor. Incoming and outgoing photon polarization along the x, y and z directions were considered. To account for finite lifetime effects a broadening with half width at half maximum (HWHM) of 0.3 eV is applied along the absorption energy axis respectively, while a broadening of 0.15 eV is applied along the loss energy axis for Fe $L_3$-edge. A global energy shift for both calculated XAS and RIXS spectra was applied to match the experimental results for individual calculations.

Charge-Transfer Multiplet calculations of XAS and RIXS spectra

The charge transfer Hamiltonian, full atomic multiplet (CTHFAM) are described in ref. (*55*). In the supplementary materials, we calculated both the Fe $L_3$ edge spectra with a single ionic atom, and hybridization with the ligand atoms. Table S2 lists all parameters used for the small cluster. Slater integrals and spin-orbit coupling values were calculated from FEFF (*56*) using the radial wavefunctions available within the FEFF10 package, with the self-consistent local density approximation. The approach is described in detail in ref. (*57*). The Slater Integrals is further reduced by 80% to fit experimental measurements. Spin orbital coupling is also calculated by averaging over the energy separation for all *jj* sub-configurations (*58*), using $d^6$ configuration for Fe and $p^4$ configuration for Sb. For ionic atoms, we show XAS and RIXS profiles in both $d^6$ and $d^7$ configurations. The low spin configuration clearly shows the lack of low energy excitations seen in the RIXS profile, ruling out the scenario of a low spin 2+ ground state. This is further supported by inspecting the Tanabe-Sugano diagrams for $d^6$ and $d^7$ configurations, where no low energy transition modes are allowed for the low spin ground state (*59*).

We also consider the case where Fe and Sb are hybridized. We model a single cluster containing 1 Fe atom and 2 Sb atoms, where the tight binding parameters for the multiplet calculations were obtained from the converged DFT ground state wavefunctions using Wannier90 (*60*), where the resulting band structure is shown in Fig. S8. Basis orbitals obtained from Wannier90 are then rotated to match the tilt of the Fe-Sb octahedral. The resulting rotated tight-binding Hamiltonian matrix elements are tabulated in table S3. The hybridization values used in the multiplet calculations are scaled proportionally from the tight-binding values obtained from Wannier90. The resulting Ground state wavefunctions are tabulated in table S2, while the orbital occupations of the ground state and the final states of interest are listed in table S4. A global energy shift is applied to all calculations to match the absorption energy observed in the experiment. The resulting XAS and RIXS show reasonable agreement with the experimental results, but does not allow us to rule out other origins for the low energy excitations.

Similar to the OCEAN calculations, the multiplet-based XAS was computed using Fermi's golden rule, while RIXS is calculated using the Kramers-Heisenberg equation (*29*). The same broadening values as OCEAN calculations were applied to both the incident energy and energy loss axis.

**Fig. S1.**

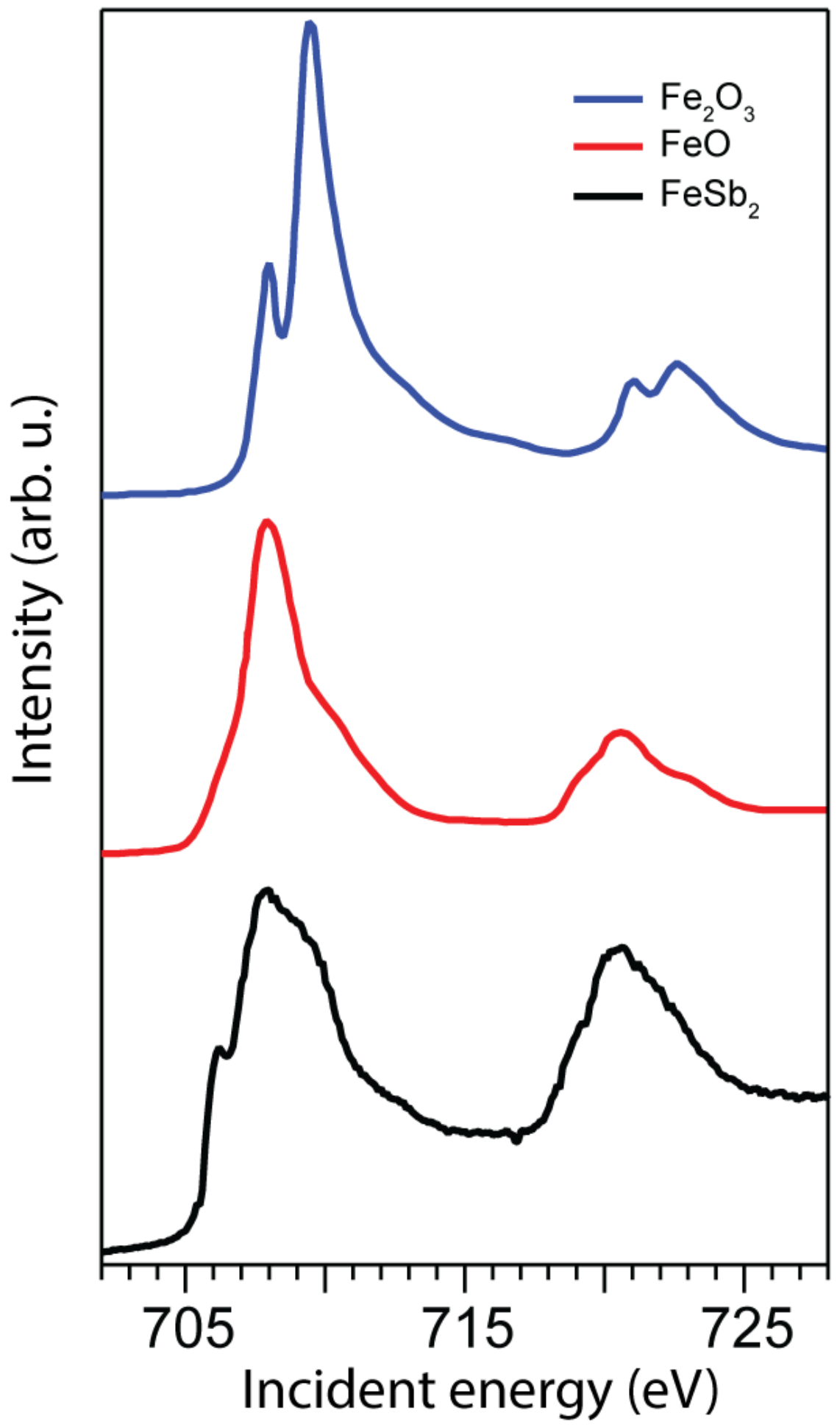

Qualitative comparison of $FeSb_2$ X-ray absorption spectroscopy (XAS) spectra to that of ionic Fe systems. The $FeSb_2$ spectra is taken with σ-polarized photons (E // *a*) at a grazing angle of 10° using total fluorescence yield. The $Fe_2O_3$ and FeO spectra are adapted from supplementary information Ref. (*61*).

**Fig. S2.**

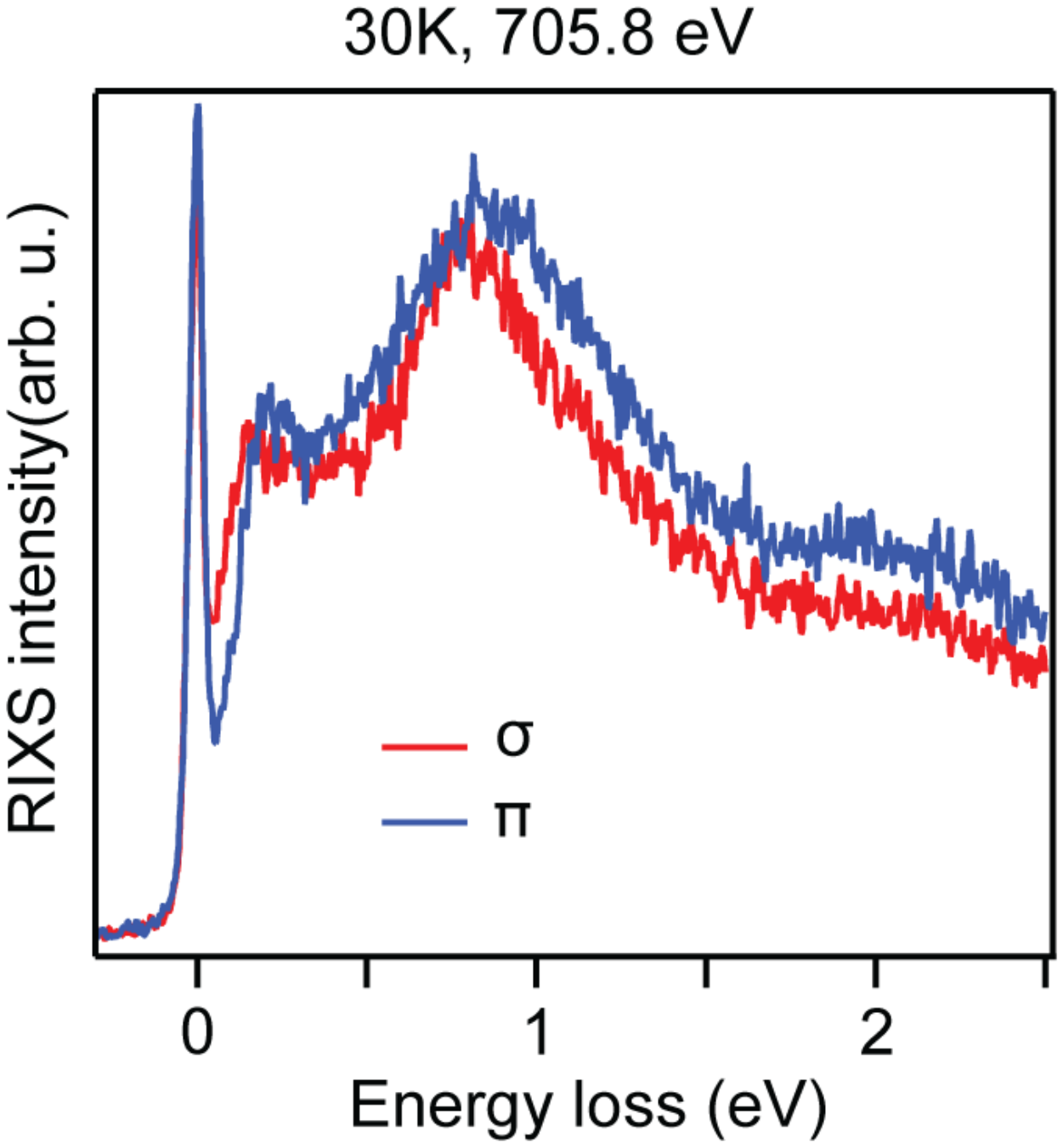

Dependence of the low energy modes at 705.8 eV incidence photon energy for different incident photon polarizations. The outgoing photon polarization is not analyzed. Here sample θ is 10 degrees, such that σ polarization corresponds to the electric field lying along the vertical *a* axis and π polarization corresponds to the electric field lying in plane, mostly along the *b* axis. Scattering arm angle 2θ is positioned at 150º.

**Fig. S3.**

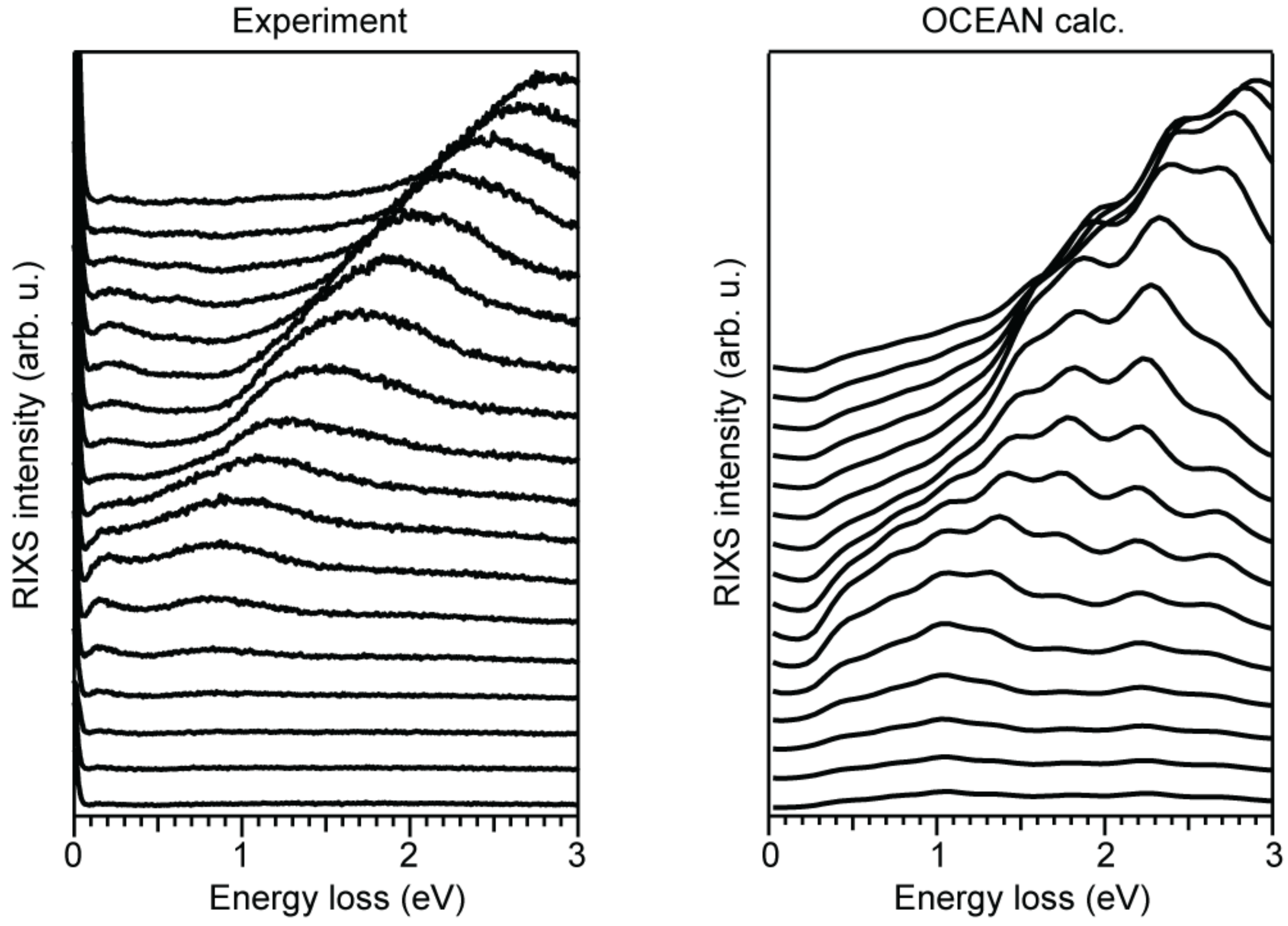

Comparison of the distribution curves between the experiment (left) and OCEAN calculations (right) for the incident energy dependent RIXS spectra. Curves are from 704.8 eV (bottom curves) to 708.2 eV (top curves) in 0.2 eV steps.

**Fig. S4.**

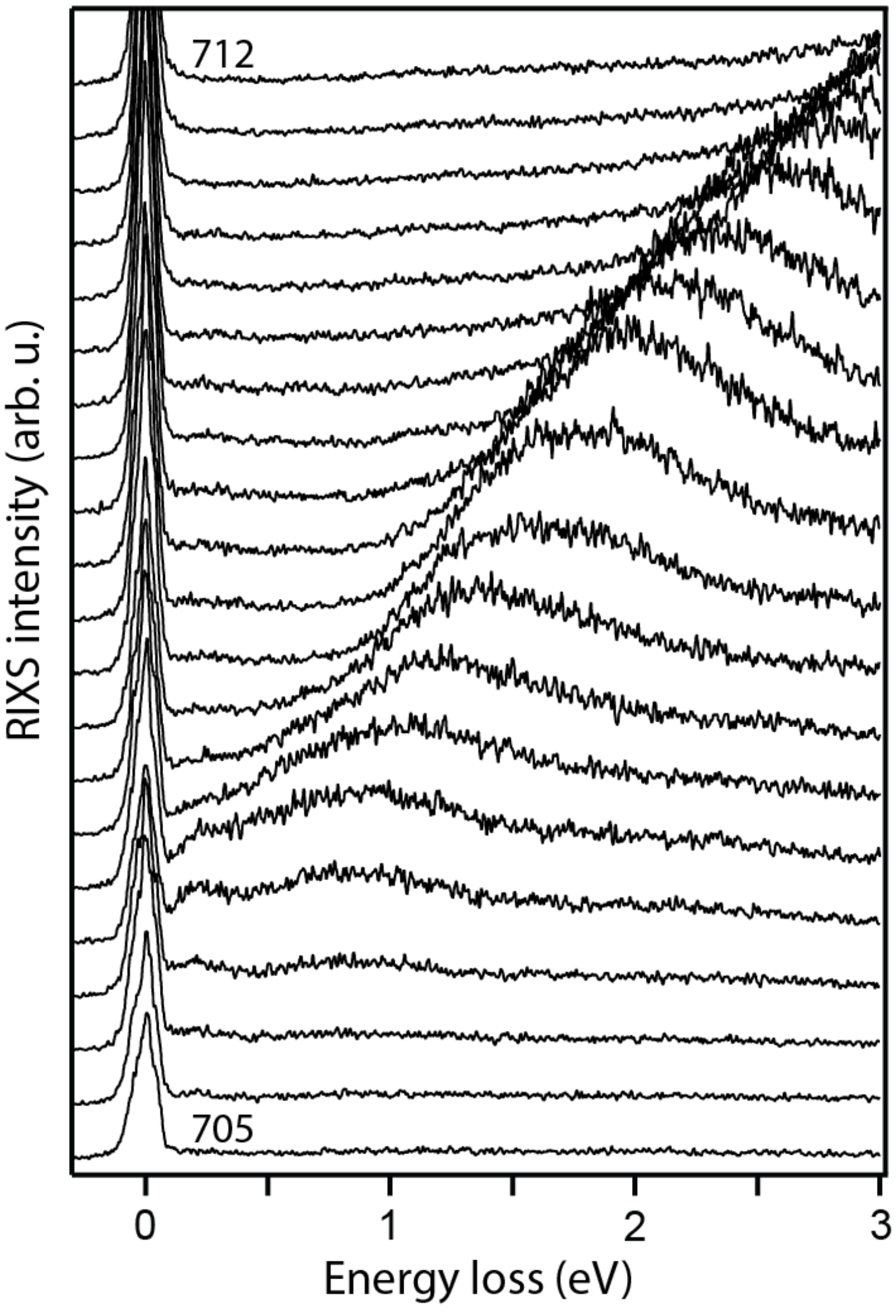

We have examined the incidence-energy-dependence of the RIXS spectra in a wider range of photon energies from 705 eV (bottom) to 712 eV (top). This covers the entire $L_3$ resonance. We do not find any noticeable Raman-like intensity at incident energies above the resonance peak of 708 eV. This indicates that the resonance profile we extracted in Fig. 1(e) reflects all of the Raman-like excitations at this edge.

**Fig. S5.**

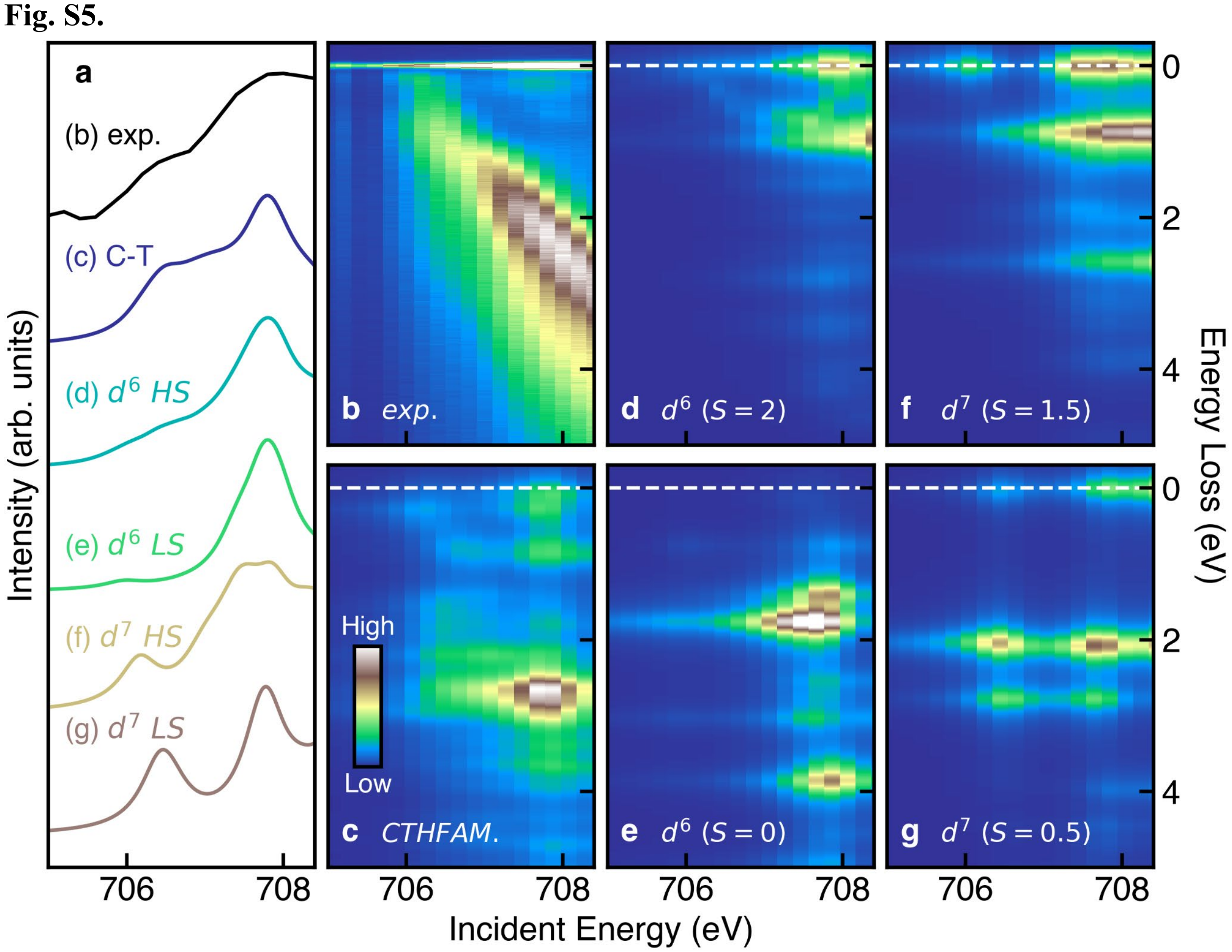

Multiplet calculations capture ground state and low energy excitations. (a) XAS spectrum comparing experiment with multiplet theory calculations. Both experiment and theory XAS curves are obtained by integrating the RIXS spectra along the energy loss axis. (b) Experimental incident-energy-dependent RIXS spectra for comparison. (c) RIXS calculated using a charge transfer Hamiltonian with full atomic multiplet (CTHFAM) that includes hybridization with Sb atoms (see methods). This model reproduces the low-energy excitations seen experimentally and provides the best overall agreement with both the XAS and RIXS profiles. While the hybridized ground state does not have good spin quantum numbers, the ground state corresponds to an S=1 configuration when spin–orbit coupling is disabled. (d) Theoretical RIXS of atomic multiplet calculation, with $d^6$ high-spin (crystal field, $\Delta_O$=1 eV, S=0), (e) $d^6$ low-spin ($\Delta_O$=2.3 eV, S=2), (f) $d^7$ high-spin ($\Delta_O$=1 eV, S=1.5) and (g) $d^7$ low-spin ($\Delta_O$=2 eV, S=0.5) configuration as a comparison to the experimental measurement. The crystal-field values for the low-spin cases are chosen at the boundary of the corresponding spin-state transitions. White dashed lines in d-g mark the zero-energy-loss elastic scattering line.

**Fig. S6.**

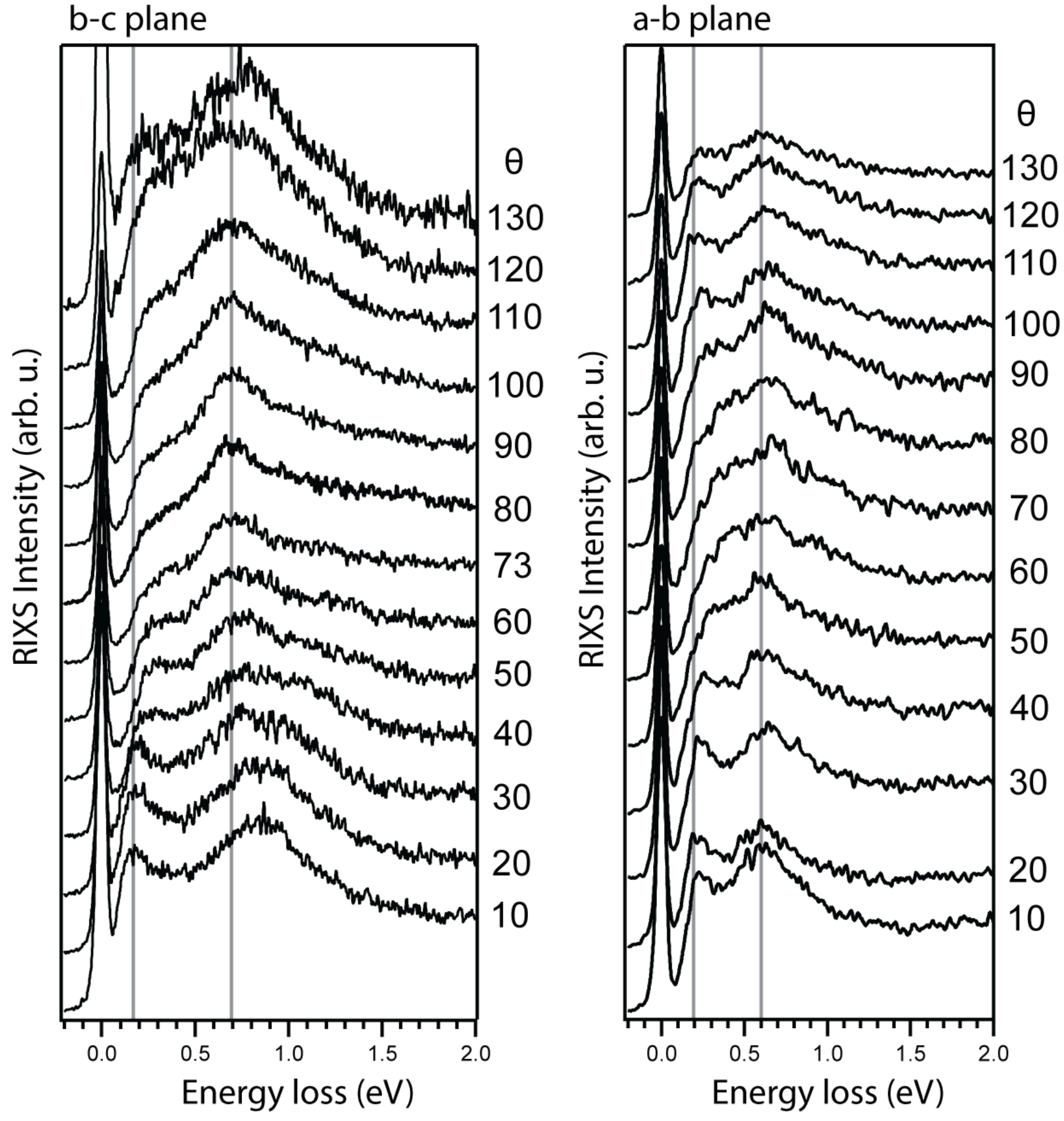

Sample θ dependence of the RIXS spectra for the *b-c* scattering plane (left) and *a-b* scattering plane (right). Numbers to the right of the curves indicate the sample θ value. The scattering arm 2θ is positioned at 150º. Grey vertical lines are energy references from the features at θ = 10º curves. Because of the apparent lack of dispersion along the *b* axis, this suggests that M1 and M2 are dispersive along the *c* axis but have little to no dispersion along the *a* axis.

**Fig. S7.**

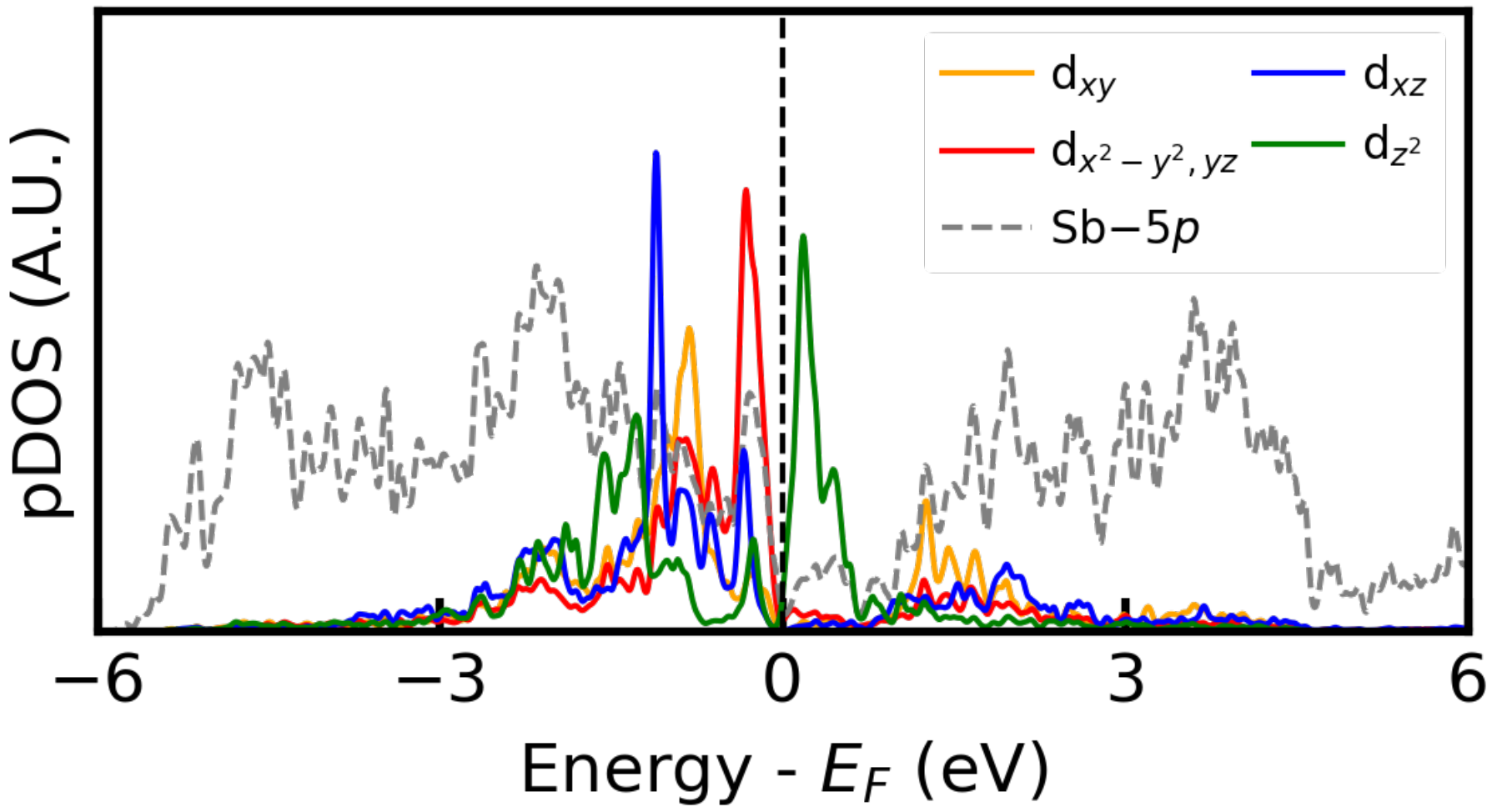

Density of states for $FeSb_2$ calculated with generalized gradient approximation (GGA) in DFT. The energy range shown here is selected for the Wannier downfolded result.

**Fig. S8.**

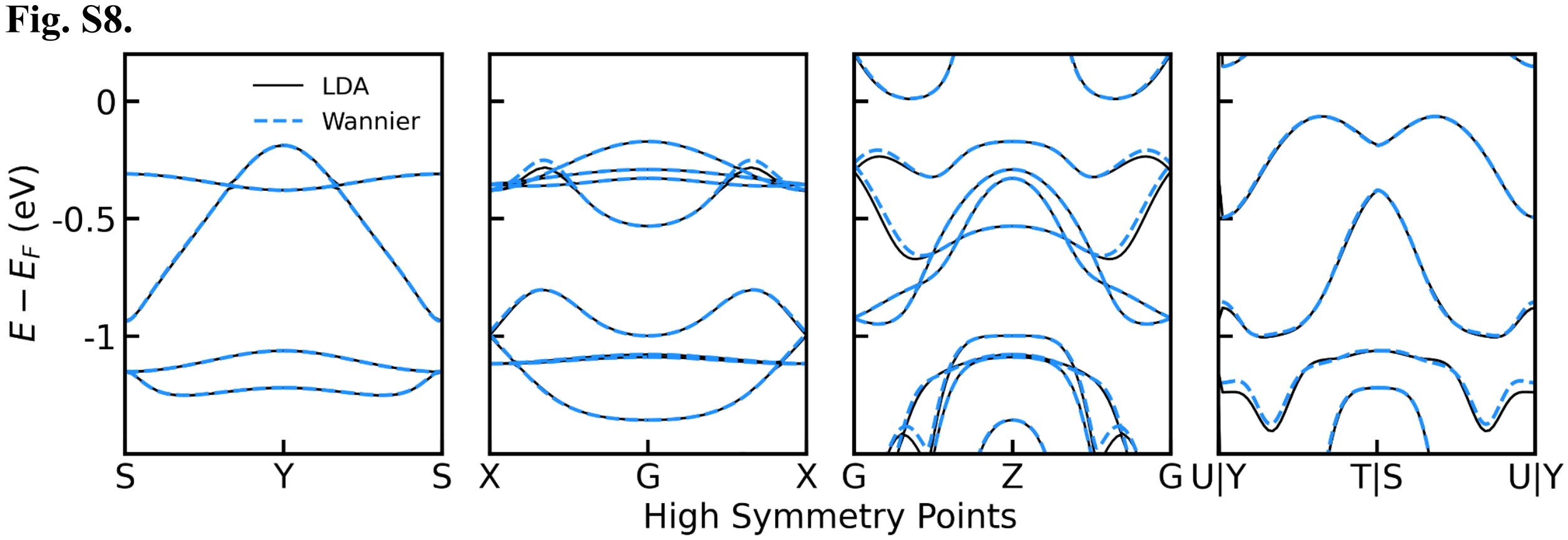

Calculated band structure of $FeSb_2$ along different high symmetry paths. The solid line is calculated by DFT, while the dashed blue line is calculated by Wannier90, downfolded from the DFT calculation in Fig. S7.

### Table S1. Bader charge for $FeSb_2$ and reference Fe compounds

Average Bader charges (referenced to the neutral Fe atoms with 16 valence electrons) of $FeSb_2$ compared to reference Fe materials. All ground state wavefunctions are calculated using the methods described in the main text. $FeSb_2$ shows nearly neutral charge on the Fe atom, indicating a highly covalent bonding environment.

| Compound | $FeSb_2$ | FeO | $Fe_2O_3$ |
|---|---|---|---|
| Fe Bader Charge | -0.15 | +1.34 | +1.67 |

## Table S2. Charge-Transfer Multiplet Calculation Parameters

Parameters used for the multiplet calculations are listed below. The single-ion multiplet and charge-transfer multiplet calculations share the same set of parameters in the row labeled **"Atomic."** Both Slater-Condon parameters for Fe ($F^2$ and $F^4$), core-valence interactions values ($F^2_{pd}$ $G^1_{pd}$ $G^3_{pd}$) and spin-orbit coupling parameters were calculated by FEFF10. The crystal-field values used for the single-ion multiplet calculations are provided in the caption of Fig. S5, whereas only the spin–orbit coupling ($\zeta$) in the core $2p$ orbital is included.

For the charge-transfer multiplet calculations, additional parameters are listed in the row labeled **"Charge Transfer."** Hybridizations ($t$) parameters are derived from tight-binding matrices obtained with Wannier90 (Table S2). $U_{dd}$, $U_{pd}$, crystal field strength ($\Delta_O$), and charge transfer energies ($\Delta$) were determined by comparison to experiment. All parameters are reported in eV. The final row provides the ground-state hole occupations used in the charge-transfer multiplet calculation.

| | **$FeSb_2$** | | | | | | | | |
|---|---|---|---|---|---|---|---|---|---|
| **Atomic** | $U_{dd}$ | $\zeta_d$ | $F^2_{dd}$ | $F^4_{dd}$ | $U_{pd}$ | $F^2_{pd}$ | $G^1_{pd}$ | $G^3_{pd}$ | $\Delta_O$ |
| | 6.0 | 9.23 | 9.21 | 5.67 | 6.0 | 5.18 | 3.82 | 2.18 | Fig. S7 |
| **Charge Transfer** | $U_{pp}$ | $\zeta_d$ | $F^2_{pp}$ | $t_{t_{2g}}$ | $t_{e_g}$ | $t_{pp}$ | $\Delta$ | $\Delta_O$ | |
| | 1.0 | 0.06 | 0.4 | 1.2 | 0.6 | 0.6 | -25 | 0.5 | |
| | Ground state hole orbital occupations for charge-transfer multiplet:<br>$\Psi_{GS} = 0.04\,\lvert d^5\underline{L}^3\rangle + 0.30\,\lvert d^6\underline{L}^4\rangle + 0.47\,\lvert d^7\underline{L}^5\rangle + 0.17\,\lvert d^8\underline{L}^6\rangle$ | | | | | | | | |

## Table S3. Tight-binding matrix element from Wannier downfolding

The procedure for obtaining the tight-binding matrix elements is described in the Methods. All values are reported in eV. These tight-binding parameters are subsequently used to inform the multiplet calculations shown in Fig. S5. We note that the strong hybridization between Fe–Sb and Sb–Sb can significantly influence the spin state of the central Fe atom.

| $FeSb_2$ | | | | | |
|---|---|---|---|---|---|
| $d_{x^2-y^2} - p_y$ | $d_{x^2-y^2} - p_x$ | $d_{xy} - p_y$ | $d_{xy} - p_x$ | $p_x - p_y$ | $p_x - p_x$ |
| 0.45 | -0.416 | -0.2889 | -0.2127 | -0.194 | 0.18 |

## Table S4. Orbital hole occupations of eigenstates for CTHFAM

Orbital occupations of different eigenstates from the CTHFAM calculation. Values are presented in hole notation, where occupation numbers represent the hole density within a specific orbital. The lowest-energy excitation in RIXS at 274 meV (indicated by the red arrow in Fig. 1f) occurs primarily between the *d* orbitals within the hybridized $e_g$-*p* molecular orbitals. The higher-energy excitation at 837 meV involves shifting hole densities from $p_{xy}/p_{yx}$ orbitals to $d_{xy}$ orbital, or equivalently, shifting electron densities from $d_{xy}$ orbital to $p_{xy}/p_{yx}$ orbitals.

| Orbital | Ground state $\vert GS\rangle$ | Final State at 274 meV $\vert M1\rangle$ | $\Delta_{\vert GS\rangle,\vert M1\rangle}$ | Final State at 837 meV $\vert M2\rangle$ | $\Delta_{\vert GS\rangle,\vert M2\rangle}$ |
|---|---|---|---|---|---|
| $d_{x^2-y^2}$ | 0.87 | 0.71 | -0.16 | 0.71 | -0.16 |
| $d_{z^2}$ | 1.37 | 1.38 | 0.01 | 1.32 | -0.05 |
| $d_{xy}$ | 0.1 | 0.88 | 0.77 | 0.76 | 0.65 |
| $d_{xz}$ | 0.42 | 0.16 | -0.26 | 0.28 | -0.14 |
| $d_{yz}$ | 0.42 | 0.16 | -0.26 | 0.28 | -0.14 |
| $p_{xx}/p_{yy}$ | 0.85 | 0.92 | 0.07 | 0.94 | 0.09 |
| $p_{xy}/p_{yx}$ | 0.97 | 1.03 | 0.06 | 0.63 | -0.34 |
| $p_{xz}/p_{yz}$ | 0.57 | 0.39 | -0.18 | 0.75 | 0.18 |